\documentstyle[12pt]{article}
\def\d{\partial}

\def\Tr{{\rm Tr}}
 
\newcommand\0{\nonumber}
\newcommand\ee{\end{eqnarray}}	 	
\newcommand\be{\begin{eqnarray}}
\newcommand\ba{\begin{array}}			
\newcommand\ea{\end{array}}

\begin{document}
\begin{flushright}
{SISSA 16/00/EP/FM}\\
{ hep-th/0002210}
\end{flushright}

\begin{center}
{\LARGE {\bf A note on consistent anomalies in}} 
\vskip 0.2cm
{\LARGE{\bf noncommutative YM theories}}
 
\vskip 1cm

{\large L. Bonora , M. Schnabl and A. Tomasiello} 
{}~\\
\quad \\
{\em International School for Advanced Studies (SISSA/ISAS),}\\
{\em Via Beirut 2, 34014 Trieste, Italy and INFN, Sezione di Trieste}\\
 {\tt bonora,schnabl,tomasiel@sissa.it}
\end{center}

\vskip 2cm
{\bf Abstract.} Via descent equations we derive formulas for consistent 
gauge anomalies in noncommutative Yang--Mills theories.

\vskip 1cm

The growing interest in noncommutative YM theories is calling our attention upon 
old problems. We would like to know if or to what extent old problems, solutions
or algorithms in commutative YM theories fit in the new noncommutative 
framework. One of these is the question of chiral anomalies .
We would like to know what form chiral anomalies take in the new noncommutative
setting. This has partially already been answered via a direct computation
\cite {Ard,GM}. In commutative YM theories there is another way to compute the 
explicit form of consistent chiral anomalies, i.e. by solving the Wess--Zumino
consistency conditions (WZcc), \cite{WZ}, in particular via the descent
equations, \cite{S,BC,ZWZ}. Now it is easy to see that in noncommutative YM theories WZcc 
still characterize chiral Ward identities, therefore
one may wonder whether one can find the form of consistent anomalies by solving
them, in particular whether a powerful algorithm like the descent equations
is still at work. The answer is yes, as we briefly illustrate below.

In this note we have in mind a YM theory in a noncommutative ${\bf R}^D$,
with Moyal deformation parameters $\theta^{ij}$. We use the form notation 
for all the expression. Therefore
we have a matrix--valued one--form gauge potential $A$, with gauge field 
strength two--form
$F = dA +A*A$. We also introduce the gauge transformation parameter $C$, in 
the form of an anticommuting Faddeev--Popov ghost (using anticommuting gauge
parameters simplifies a lot anomaly formulas, as is well--known).
Next we introduce the following gauge transformation conventions:
\be
\delta A = dC + A*C-C*A,\quad\quad \delta C = -C*C\label{brst}
\ee
The differential $d$ is defined in the general non--commutative geometry 
setting as follows: it is the exterior derivative
of the universal differential graded algebra $\Omega({\cal B})$ associated to 
any algebra ${\cal B}$ \cite{connes,madore,landi}; ${\cal B}$ is for us the algebra
generated by $A$ and $C$. In simple words, this means that we deal with forms 
as usual, but never use the relation: $ \omega_1\omega_2
= (-)^{k_1k_2}\omega_2\omega_1$, for any $k_i$--form $\omega_i$.

$d$ and $\delta$ are assumed to commute. As a consequence
the transformations (\ref{brst}) are nilpotent as in the commutative case. 
They are noncommutative BRST transformations.
 
If one tries to derive for commutative YM theories descent equations similar to 
those of the commutative case, at first sight this seems to be impossible.
In fact the standard expression one starts with, $\Tr(F\ldots F)$, in the 
commutative case should be replaced by  $\Tr(F*\ldots *F)$ ($\Tr$ denotes 
throughout the paper the trace over matrix indices\footnote{On more general
noncommutative spaces (other than ${\bf R}^D$) this trace without an
accompanying integration may not be well defined.} ; but the latter is neither 
closed nor invariant, as one may easily realize. However one notices that 
it would be both closed and invariant if we were allowed to permute cyclically 
the terms under the trace symbol. In fact, terms differing by a cyclic 
permutation differ by a total derivative of the form $\theta^{ij}\d_i...$.
Such terms could of course be discarded upon integration. However, the 
spirit of the descent equations requires precisely to work with unintegrated 
objects.
 
The way out is then to define a bi--complex which does the right job. It is 
defined as follows. Consider the space of (${\cal A}$--valued, where ${\cal A}$
is the algebra defining our non--commutative space)
traces of $\ast$ products of such objects as $A,dA,C,dC$.   
The space of cochains is now this space, modulo the circular relation
\be
\Tr(E_1*E_2*...*E_n)\approx \Tr(E_n*E_1*...*E_{n-1}) (-1)^{k_n(k_1+\ldots+k_{n-1})}
\label{circ}
\ee
where $E_i$ is any of $A,dA,C,dC$, and $k_i$ is the order form of $E_i$.

The definition of the bi--complex, let us call it ${\cal C}$, is completed
by introducing two differential operators.
The first is $d$, as defined above. 
The second differential is $\delta$, the BRST cohomology operator.
We define it to commute with $d$. Both preserve the relation (\ref{circ}).

We can now start the usual machinery of consistent anomalies, reducing the
problem to a cohomological one. 
In a noncommutative even D--dimensional space we start with
$\Tr(F*F*...*F)$ with $n$ entries, $n= D/2+1$. In the complex ${\cal C}$ this
expression is closed and BRST--invariant. Then it is easy to prove 
the descent equations:
\be
&&\Tr(F*F*...*F)= d \Omega_{2n+1}^0\0\\
&&\delta \Omega_{2n+1}^0= d \Omega^1_{2n}\label{descent}\\
&&\delta \Omega^1_{2n} = d \Omega^2_{2n-1}\0
\ee
and so on. Here the Chern--Simons term can be represented in ${\cal C}$ by
\be
\Omega_{2n+1}^0 = n \int_0^1 dt \Tr (A*F_t*F_t*...*F_t)\label{CS}
\ee
where we have introduced a parameter $t$, $0\leq t\leq 1$, and 
the traditional notation $F_t= tdA +t^2A*A$.

The anomaly can instead be represented by
\be
\Omega^1_{2n} &=& n\int_0^1 dt (t-1)\Tr(dC*A*F_t*...*F_t+dC*F_t*A*...*F_t+\0\\
&&\ldots+dC*F_t*F_t*...*A) 
 \label{anom}
\ee
where the sum under the trace symbol includes $n-1$ terms.  

Finally
\be
\Omega^2_{2n-1} = n\int_0^1 dt \frac {(t-1)^2}{2} \Tr (dC*dC*A*F_t*...*F_t+
\ldots)\0
\ee
where the dots represent $(n-1)(n-2)-1$ terms obtained from the first 
by permuting in all 
distinct ways $dC, A$ and $F_t$, keeping track of the grading and keeping
$dC$ fixed in the first position. 

The only trick to be used in proving 
the above formulas is to assemble terms in such a way as to form the 
combination $dA +2t A*A= \frac{dF_t}{dt}$, and then integrate by parts.

In four dimensions the anomaly takes the form
\be
\Omega_4^1= -\frac 12 \Tr(dC*A*dA+ dC*dA*A + dC*A*A*A)\label{4dim}
\ee
This anomaly, once it is integrated over, coincides with the result of 
\cite{GM}, eq.(24) (modulo conventions).

On the basis of the above exercise, noncommutativity exhibits new qualitative
features for chiral anomalies. Let us consider the case of 
$A= \sum_a A^a T^a$, where $T^a$ is 
a basis of antihermitean matrices. The first two terms in (\ref{4dim}) are 
proportional to $\Tr(T^aT^bT^c)$. Therefore the anomaly (\ref{4dim}) vanishes
only if $\Tr(T^aT^bT^c)=0$, as was noticed in \cite{GM}. Now, 
$\Tr(T^aT^bT^c)= \frac 12 \Tr(T^a\{T^b,T^c\}) +\frac 12 \Tr(T^a[T^b,T^c])$.
The first term in the RHS is the usual ad--invariant third order tensor;
the second term, which is absent in the commutative case, is proportional 
to the structure constant and vanishes only when
all the structure constants do. Analogous arguments apply to other dimensions.
However the existence of a new part of the chiral anomaly that vanishes in the
commutative case is of no use in the analysis of possible new
cancellation mechanisms as long as the gauge group is $U(N)$, because 
the cancellation is driven by the $U(1)$ factor. 
In this case we reach the conclusion that noncommutative anomalies
cannot vanish due to vanishing of ad--invariant
tensors, as it occurs for many gauge groups in the commutative case:
the only vanishing mechanism is therefore the one  
produced by matching anomaly coefficients with opposite chirality.

\vskip .3cm
{\bf Note} After this work was completed, J.Mickelsson informed us that the
descent equations for chiral anomalies have been previously studied in
\cite{lang}. The cohomology used in the two papers are different. Moreover
the method used here is so much simpler, with results spelled out in detail,
that we deem it worth a short note.

\vskip .5cm
{\bf Acknowledgements}
We would like to thank C.P.Martin for a useful exchange of e-mail messages
and L.Dabrowski, T. Krajewski and G.Landi for useful discussions.
We thank J.Mickelsson for pointing out to us ref.\cite{lang}. 
This research was partially supported by EC TMR Programme,
grant FMRX-CT96-0012, and by the Italian MURST for the program
``Fisica Teorica delle Interazioni Fondamentali''.

\end{document}